# Integrated magnonic neural circuits based on nonlinear wave neurons

*Mengying Guo[1*], Xudong Jing[1*], Kristýna Davídková[2*], Roman Verba[3], Zhenyu Zhou[1], Xueyu Guo[1], Carsten Dubs[4], Chuan Gao[5], Yiheng Rao[6], Kaiming Cai[1], Jing Li[7], Philipp Pirro[8], Andrii V. Chumak[2], Qi Wang[1†]*

*[1] School of Physics, Hubei Key Laboratory of Gravitation and Quantum Physics, Institute for Quantum Science and Engineering, Huazhong University of Science and Technology, Wuhan, China*

*[2] Faculty of Physics, University of Vienna, Vienna, Austria*

*[3] V. G. Baryakhtar Institute of Magnetism of the NAS of Ukraine, Kyiv, Ukraine*

*[4] INNOVENT e.V., Technologieentwicklung, Jena, Germany*

*[5] Wuhan National Laboratory for Optoelectronics, Huazhong University of Science and Technology，Wuhan，China*

*[6] School of Microelectronics, Hubei University, Wuhan, China*

*[7] Wuhan National High Magnetic Field Center, Huazhong University of Science and Technology, Wuhan, China*

*[8] Fachbereich Physik and Landesforschungszentrum OPTIMAS, Rheinland-Pfälzische Technische Universität Kaiserslautern-Landau, Kaiserslautern, Germany*

Artificial intelligence is driving intense interest in alternative computing hardware capable of neural information processing beyond conventional charge-based electronics. Among emerging approaches, wave-based computing promises highly parallel and energy-efficient operation, but scalable physical neural hardware has remained elusive because wave systems generally lack cascadable nonlinear neurons with signal regeneration and phase-robust operation. Here we demonstrate integrated magnonic neural circuits based on nonlinear threshold neurons realized in nanoscale yttrium iron garnet waveguides. The neurons perform weighted summation of multiple spin-wave inputs, while a pump-controlled nonlinear activation defines continuously tunable firing thresholds. Owing to deeply nonlinear spin-wave dynamics, the activated neurons emit self-normalized outputs whose intensities are largely independent of the input amplitudes, while nonlinear phase self-adjustment suppresses sensitivity to the relative input phases, enabling deterministic neuron-to-neuron cascading without external signal restoration. We experimentally realize programmable threshold neurons, reconfigurable weighted classification and deterministic cascading between sequential neuronal stages, and further demonstrate reconfigurable physical pattern recognition in a seven-neuron integrated magnonic circuit through experimental classification of the binary letter patterns “HUST”. These results establish nonlinear magnons as a scalable platform for integrated neural hardware and position nonlinear wave dynamics as a general paradigm for physical neuromorphic computing.

[*] These authors contributed equally to this work.

[†] Corresponding Author: williamqiwang@hust.edu.cn

## Introduction

The rapid expansion of artificial intelligence has created an urgent demand for alternative computing hardware capable of overcoming the energy consumption and architectural limitations of conventional charge-based electronics. Although modern artificial neural networks achieve remarkable performance in tasks such as pattern recognition, language processing and scientific discovery, their implementation on CMOS hardware remains fundamentally constrained by the von Neumann bottleneck, data-transfer overhead and the increasing power cost of large-scale neural computation. These challenges have stimulated intense interest in neuromorphic and wave-based computing platforms that can process information through collective physical dynamics rather than sequential charge transport [1-6]. In particular, wave systems provide a natural framework for parallel summation, interference and nonlinear activation, offering attractive opportunities for highly efficient neural information processing. However, cascadable nonlinear neurons with signal regeneration and robustness against phase fluctuations are a limiting bottleneck to construct more efficient scalable neural hardware.

Among emerging wave-computing approaches, magnonics - which utilizes spin waves as information carriers - provides several unique advantages, including nanoscale wavelengths, intrinsic nonlinearity, low dissipation and direct compatibility with microwave electronics [7-11]. Recent years have witnessed rapid progress in magnonic devices, including spin-wave logic [12-18], inverse-designed functional elements [19-21], nonlinear magnonic systems [21-23], and wave-based neural concepts [23,24]. These advances establish magnons as a promising platform for compact and energy-efficient wave computing. However, most existing magnonic implementations remain isolated proof-of-concept devices rather than scalable neural hardware architectures.

This challenge originates from the intrinsic properties of wave-based information processing. In many previous magnonic logic and majority-gate schemes, information is encoded in the phase of spin waves [25-30]. Consequently, the output strongly depends on propagation phase accumulation, fabrication imperfections and device geometry, making reliable cascading extremely difficult in large-scale circuits. At the same time, magnetic damping continuously attenuates propagating spin waves, so that the output intensity of one device generally cannot directly drive the next without additional amplification or conditioning [14, 31-34]. Furthermore, most demonstrated magnonic devices lack programmable weighting and tunable threshold activation, which are fundamental building blocks of neural computation. Overcoming these limitations requires a fundamentally different operating principle in which nonlinear wave dynamics intrinsically provide thresholding, normalization and phase-robust signal transfer. These capabilities also remain a major challenge in integrated photonic systems [35, 36].

Here we demonstrate integrated magnonic neural circuits based on nonlinear threshold neurons realized in nanoscale yttrium iron garnet (YIG) waveguides. The neuron performs weighted summation of multiple spin-wave inputs, while a pump-controlled nonlinear activation defines continuously tunable firing thresholds. Owing to deeply nonlinear spin-wave dynamics, the activated neuron emits self-normalized outputs whose intensities are largely independent of the input amplitudes, while nonlinear phase self-adjustment suppresses sensitivity to the relative phases of the incoming spin waves. These properties enable deterministic neuron-to-neuron cascading and establish nonlinear magnons as functional building blocks for scalable wave-native neural hardware.

Using micro-focused Brillouin light scattering (μBLS) spectroscopy, we experimentally demonstrate programmable threshold neurons, reconfigurable weighted classification and deterministic cascading between sequential neuronal stages. Finally, we fabricate and operate a multi-neuron integrated magnonic chip containing seven interconnected neurons, and experimentally demonstrate neural-network-based recognition of the letters "HUST". Together, these results establish a scalable wave-based neural hardware platform and provide a pathway towards integrated magnonic neural networks for energy-efficient information processing.

**Results**

Figure 1a illustrates a biological neuron, in which electrical signals from pre-neurons are integrated within the cell body and trigger neuronal firing once the total input exceeds a threshold. This process can be simplified into the artificial-neuron model shown in Fig. 1b, consisting of weighted inputs, signal summation and a nonlinear activation function. Based on this principle, we fabricate a nonlinear magnonic threshold neuron using propagating spin waves in nanoscale YIG waveguides, as shown in Fig. 1c.

The three-input magnonic threshold neuron is fabricated from a 47-nm-thick YIG film [37,38] using electron-beam lithography, in which a positive electron-beam resist was employed directly as the etching mask during ion-beam milling (see Methods). The device consists of three 800-nm-wide YIG waveguides connected to a central combining region and a pump-controlled output waveguide. Four gold stripe antennas are patterned on top: three as input ports and one that acts as a pump controlling the nonlinear activation of the device. As indicated in Fig. 1c, the antenna widths are 1.4 μm and 1.6 μm, respectively. An external magnetic field of 320 mT is applied along the z-axis to saturate the magnetization out-of-plane, allowing the use of forward volume spin waves that support a strong nonlinear behaviour [34,39,40]. Microwave pulses (0.8 μs duration, 1 μs period) at 5.1 GHz are applied to both the input and the pump antennas. The microwave power is adjusted so that spin waves are directly excited at the inputs, while the pump antenna is operated within its bistable region, where the magnon state (low or high intensity) depends on its excitation history [39-42]. At the selected operating condition, the pump power alone remains below the direct spin-wave excitation threshold and therefore cannot independently generate propagating spin waves [43,44] [see Supplementary Materials].

When spin waves from multiple inputs propagate into the combining region beneath the pump antenna, they collectively modify the local magnonic state. Once the total incoming spin-wave intensity exceeds a pump-dependent nonlinear threshold, the pump antenna switches into a high-emission state and re-emits spin waves into the output waveguide. This nonlinear re-emission process plays the role of neuronal firing. Importantly, owing to the deeply nonlinear excitation mechanism, the emitted output signal becomes self-normalized and largely independent of the exact input intensity once the activation threshold is exceeded [40]. The nonlinear neuron therefore simultaneously provides threshold activation and intrinsic signal regeneration.

The neuronal activation threshold can be continuously controlled through the pump power, as summarized in Fig. 1d. When the pump power lies within the gray region, the activation threshold is high and all three inputs are required simultaneously to trigger the pump antenna (N=3). Increasing the pump power progressively reduces the activation threshold, such that the neuron switches successively into the N=2 and N=1 regimes, where two or one active inputs are sufficient to trigger neuronal firing, respectively. At still higher pump powers, the pump antenna directly excites spin waves even in the absence of external inputs (N=0). Physically, this behavior originates from the power-dependent shift of the low-frequency edge of the bistable region toward the operating frequency. Increasing the pump power reduces the frequency gap between the operating frequency and the bistable edge, thereby moving the system closer to the nonlinear switching condition [see Supplementary Materials]. These results demonstrate continuous pump-power programmability of the neuronal activation threshold within the same physical architecture.

In addition to threshold programmability, the nonlinear magnonic neuron also supports continuously tunable input weights. As shown in [45], a direct current applied to the input antenna generates an asymmetric Oersted field near the antenna edges, partially reflecting the propagating spin waves and thereby modifying the transmission efficiency of the corresponding input channel, see Fig. 1e. As a result, the effective neuronal weight can be continuously tuned from nearly 1 to 0. This mechanism enables current-controlled reconfiguration of the weighting function within the integrated magnonic neural circuit.

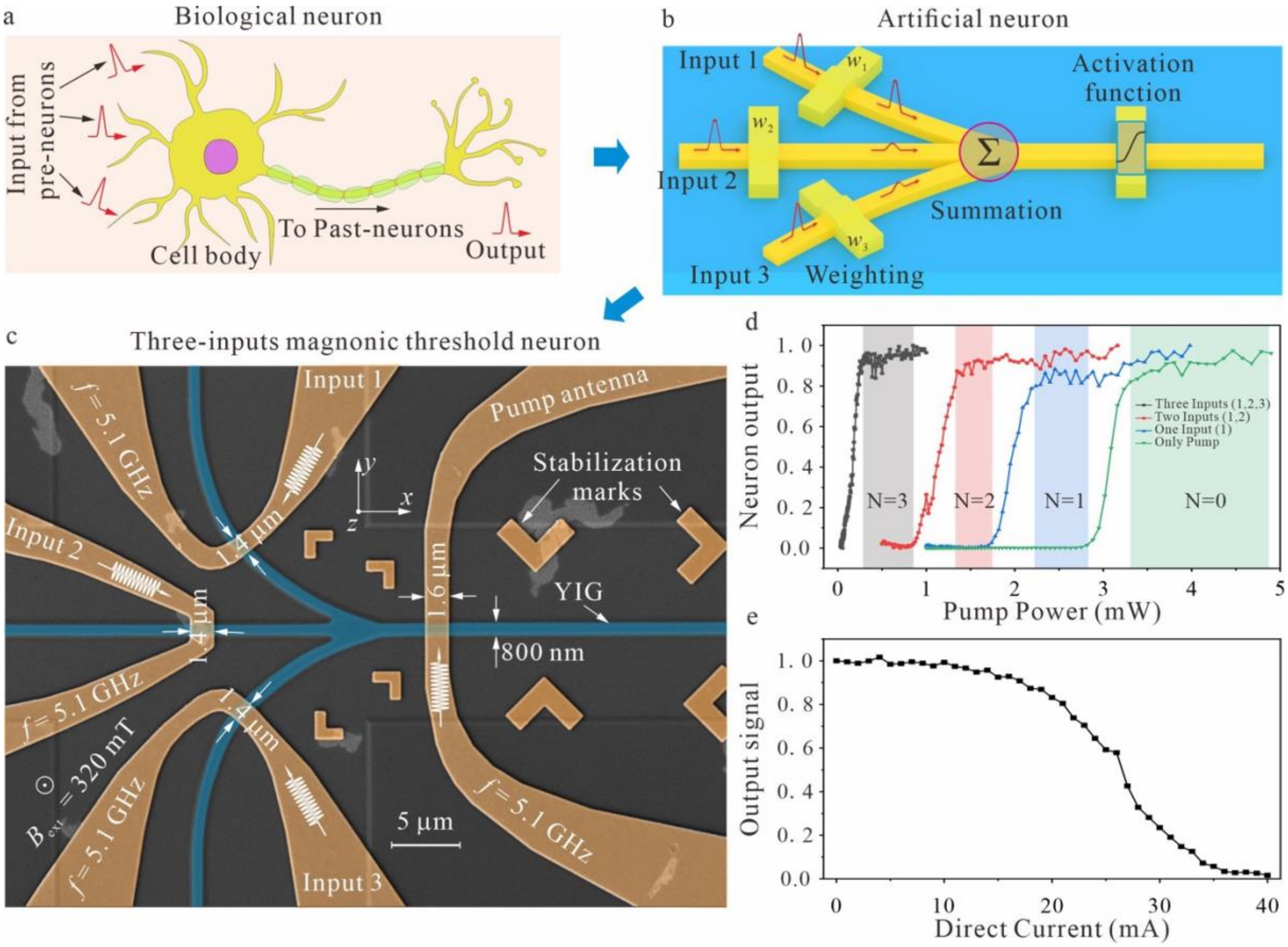

*Figure 1* ***Concept and experimental realization of a nonlinear magnonic threshold neuron. a,*** *Schematic illustration of a biological neuron. Electrical signals from pre-neurons are integrated within the cell body, and once the total input exceeds a firing threshold, the neuron transmits an output signal to subsequent neurons.* ***b,*** *Simplified artificial-neuron model consisting of weighted inputs ($w_1$, $w_2$, $w_3$), signal summation and a nonlinear activation.* ***c,*** *Scanning electron microscope (SEM) image of the experimentally realized three-input nonlinear magnonic threshold neuron fabricated from a 47-nm-thick YIG film. Three input waveguides merge into a central combining region connected to a pump-controlled output waveguide. Gold stripe antennas are used to excite propagating spin waves at the inputs and to define the nonlinear activation region. An out-of-plane magnetic field of 320 mT establishes the forward-volume spin-wave geometry.* ***d,*** *Continuously tunable neuronal activation threshold controlled by the pump power. The normalized neuronal output is plotted as a function of pump power for different input combinations. In the gray region, all three inputs are required to activate the neuron (N=3); in the pink region, any two inputs are sufficient (N=2); in the blue region, a single input can activate the neuron (N=1); and in the green region, the pump antenna directly excites spin waves without external inputs (N=0).* ***e,*** *Electrical control of the neuronal input weight. The normalized output signal is plotted as a function of direct current applied to the input antenna. The asymmetric Oersted field generated by the direct current partially or completely reflects propagating spin waves, enabling continuous tuning of the effective neuronal weight from nearly 1 to 0.*

Building on the programmable threshold activation described above, information is encoded in the spin-wave intensity, with low and high intensity representing states ‘0’ and state ‘1’, respectively. When the neuronal activation threshold is set to the N=2 regime, the three-input magnonic neuron naturally performs a majority operation, generating a high-output state only when at least two inputs are simultaneously active. Under this condition, propagating spin waves from multiple inputs collectively exceed the nonlinear activation threshold of the pump antenna, triggering neuronal firing through nonlinear spin-wave re-emission into the output waveguide.

Figure 2a-h presents two-dimensional μBLS intensity maps [46] measured for all possible input configurations of the three-input magnonic neuron. The measurements were obtained by scanning the laser spot across the center of the device over a 15×8.7 $\mu m^2$ area with a resolution of 40×20 points and integrating the spin-wave intensity within the frequency range from 4.8 to 5.6 GHz. For the input states “110”, “101”, “011” and “111” (Fig. 2a-d), the combined spin-wave intensity in the central combining region exceeds the nonlinear activation threshold, resulting in strong spin-wave emission into the output waveguide corresponding to the neuronal high-output state. In contrast, for the input states “100”, “010”, “001” and “000” (Fig. 2e-h), the incoming spin-wave intensity remains below threshold, preventing activation of the nonlinear emission state and leaving the neuron in the low-output regime.

The line profiles shown on the right side of each panel quantify the spin-wave intensity integrated within the red dashed rectangular region near the output waveguide. The measurements clearly reveal two well-separated output states corresponding to neuronal firing and non-firing conditions. Importantly, once the nonlinear threshold is exceeded, the emitted output intensity becomes nearly independent of the exact input configuration owing to the self-normalized nonlinear emission process. Figure 2i presents a reference measurement with the pump antenna switched off, confirming that the strong output signal originates predominantly from the nonlinear re-emission of the pump antenna rather than direct propagation of the attenuated input spin waves. The nonlinear magnonic neuron therefore simultaneously provides threshold activation and intrinsic signal regeneration, while majority functionality naturally emerges as one operating regime of the programmable threshold neuron.

Unlike conventional phase-encoded magnonic majority gates based on linear wave interference [30,47], the present neuron operates through intensity-based threshold activation. Consequently, the output is determined primarily by the total spin-wave intensity rather than by the relative phases of the input waves, substantially improving robustness against phase fluctuations and fabrication-induced propagation variations. The physical origin of this phase robustness is investigated in detail later. Importantly, the majority functionality is not permanently defined by the device geometry, but instead emerges from the programmable threshold behavior of the nonlinear magnonic neuron. By tuning the neuronal threshold, the same physical architecture can be dynamically reconfigured to perform different computational functions.

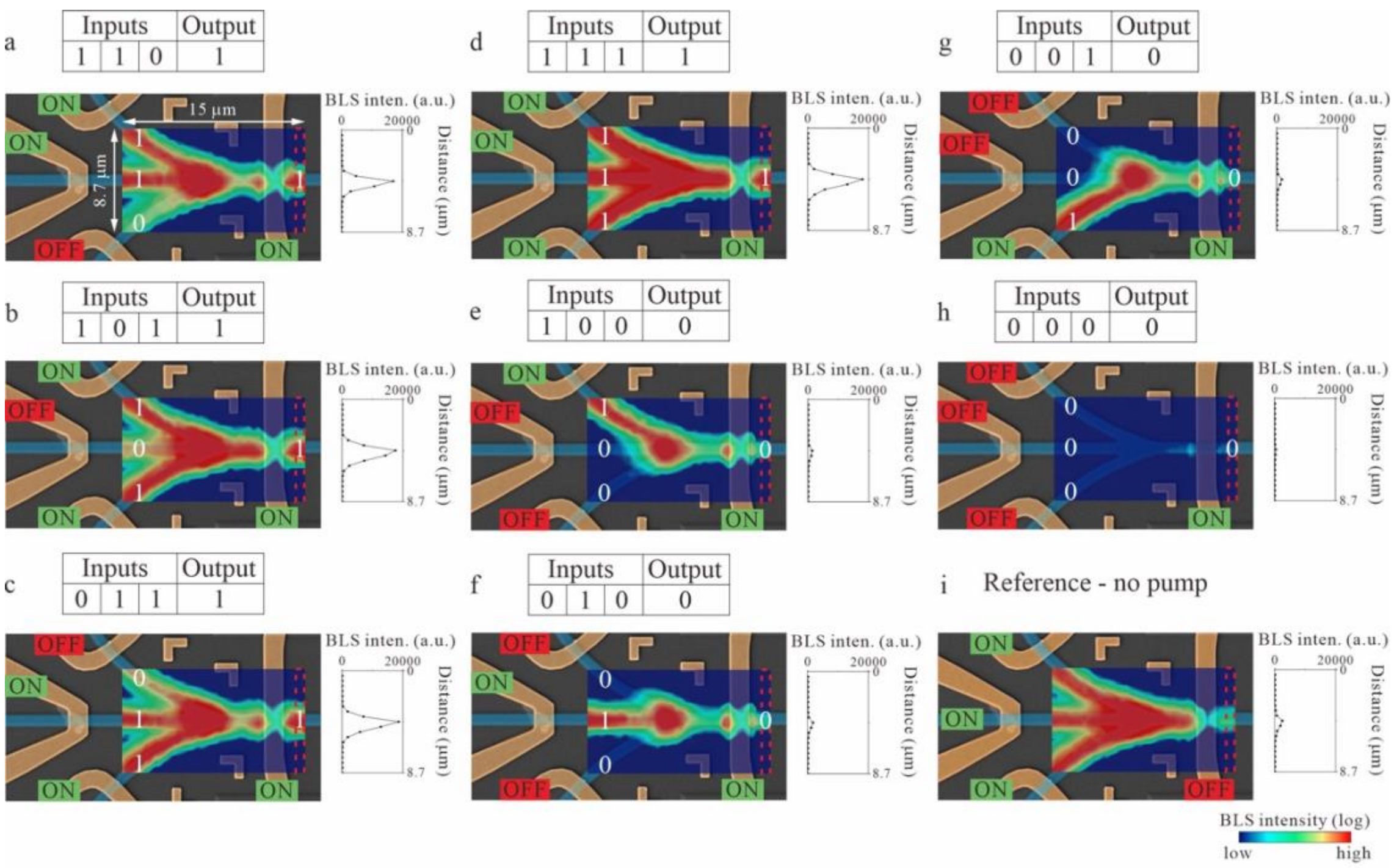

*Figure 2* ***Three-input nonlinear magnonic threshold neuron operating in the majority regime.*** *Two-dimensional μBLS intensity maps measured for different input configurations of the three-input nonlinear magnonic neuron. The activation threshold of the pump antenna is tuned such that the neuron fires only when at least two inputs are simultaneously active (N=2).* ***a–d,*** *Input states "110", "101", "011" and "111", for which the combined spin-wave intensity exceeds the nonlinear activation threshold, triggering strong nonlinear spin-wave re-emission into the output waveguide corresponding to the neuronal high-output state.* ***e–h,*** *Input states "100", "010", "001" and "000", for which the total incoming spin-wave intensity remains below threshold and the neuron stays in the low-emission state. The right panels show the spin-wave intensity integrated over the red dashed rectangular regions near the output waveguide.* ***i,*** *Reference measurement with the pump antenna switched off, confirming that the high-output state originates predominantly from nonlinear re-emission of the pump antenna rather than direct propagation of the attenuated input spin waves. Once the nonlinear activation threshold is exceeded, the emitted output intensity becomes nearly independent of the exact input configuration owing to the self-normalized nonlinear excitation process.*

Having established programmable threshold activation, we next show that the magnonic neuron can further perform weighted classification through electrical control of the individual input channels. Similar to biological synapses and artificial neural networks, different inputs contribute unequally to the neuronal activation process. In the present system, this functionality is achieved by applying direct currents to selected input antennas, thereby modifying the propagation efficiency of spin waves through the locally generated asymmetric Oersted fields.

Figure 3 summarizes the weighted classification capability of the three-input nonlinear magnonic neuron operating in the N=2 regime. The neuronal output is measured for different input combinations while selectively tuning the effective weights of individual input channels. In Fig. 3a, all three inputs are assigned equal weights ($w_1 = w_2 = w_3 = 1$). Under this condition, the neuron implements a majority-gate function, generating a strong output whenever at least two inputs are simultaneously activated. Consequently, the input states "110", "101" and "011" all produce similarly high neuronal outputs, while the remaining configurations remain below threshold.

The neuronal response can be dynamically reconfigured by electrically suppressing selected input channels. In Fig. 3b, the weight of Input 2 is reduced to nearly zero through direct-current control ($w_1 = w_3 = 1, w_2 = 0$).

As a result, only the input pattern "101", in which the two weighted channels are simultaneously activated, generates a strong neuronal output. All other input combinations remain below the activation threshold, even when two physical inputs are present. Similarly, in Fig. 3c, the weight configuration is adjusted to $w_1 = w_2 = 1$ and $w_3 = 0$. Under this case, the neuron selectively responds to the input pattern "110", while the remaining configurations produce only weak outputs.

These results demonstrate that the same physical neuron can be dynamically reconfigured to perform different classification tasks without modifying the device geometry. Importantly, the weighting operation is achieved entirely through electrical control of spin-wave propagation, enabling continuous and programmable adaptation of the neuronal response within the integrated magnonic circuit. Unlike conventional fixed-function magnonic logic gates, the present architecture therefore operates as a reconfigurable nonlinear classifier analogous to a perceptron neuron in artificial neural networks.

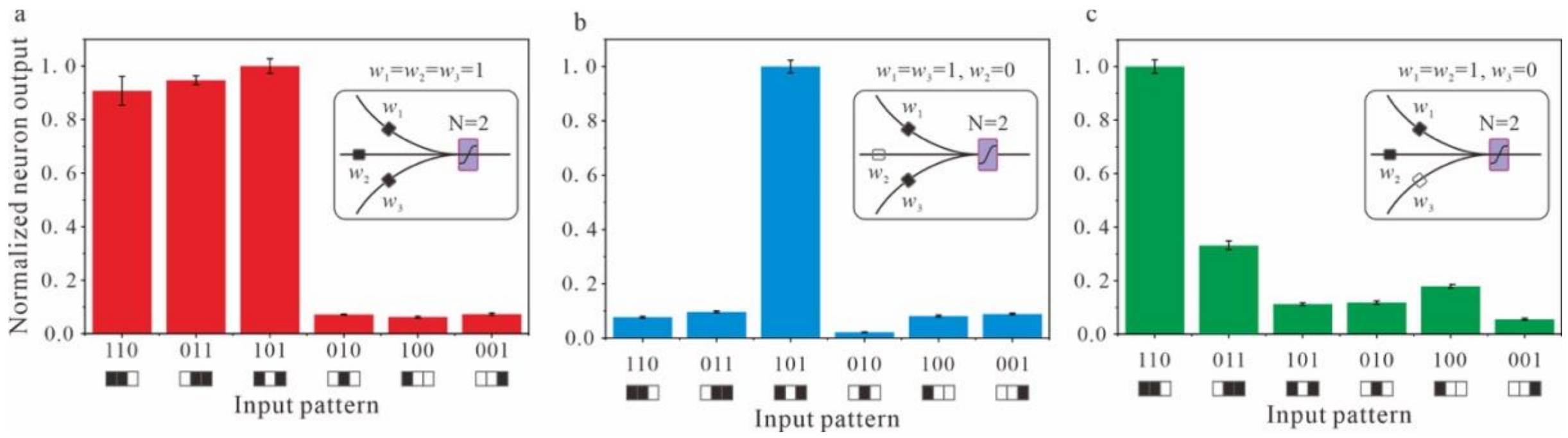

*Figure 3* ***Electrically programmable weighted pattern recognition using a nonlinear magnonic threshold neuron.*** *Normalized neuronal output measured for different input patterns under electrically controlled weighting conditions. The neuron operates in the N=2 threshold regime.* ***a,*** *Equal-weight configuration with $w_1 = w_2 = w_3 = 1$. Under this condition, the neuron operates in the majority regime and generates strong outputs for the input patterns "110", "101" and "011".* ***b,*** *Weighted configuration with $w_1 = w_3 = 1$ and $w_2 = 0$, realized by electrically suppressing Input 2 through direct-current control. The neuron selectively recognizes the input pattern "101", while all other patterns remain below threshold.* ***c,*** *Weighted configuration with $w_1 = w_2 = 1$ and $w_3 = 0$, where the neuron selectively responds to the input pattern "110". Insets illustrate the corresponding weighting configurations for each measurement. Error bars represent the standard deviation obtained from three repeated measurements.*

The weighted pattern-recognition capability demonstrated above establishes the proposed device as a programmable nonlinear magnonic neuron. However, scalable neural hardware further requires robust signal propagation between neuronal stages, including immunity to phase fluctuations, intrinsic signal regeneration and deterministic cascading. In conventional interference-based computing systems, the output strongly depends on the relative phases and amplitudes of the incoming waves, making reliable multi-stage integration extremely challenging. Small fabrication imperfections and propagation inhomogeneities continuously accumulate phase errors during wave propagation, making precise phase control increasingly difficult in large-scale integrated circuits. The nonlinear magnonic neuron intrinsically overcomes these limitations through nonlinear phase self-adjustment and self-normalized signal regeneration.

To directly investigate the reason of the phase-robustness of the neuronal response, we fabricated a simplified Y-shaped magnonic structure consisting of two input waveguides and one pump-controlled output branch. Figure 4a summarizes the measured output intensity as a function of the relative phase difference between the two input spin waves. Without the pump activation, the detected output remains weak because only directly transmitted spin waves reach the output region. In contrast, once the pump antenna is activated, the output intensity increases by nearly a factor of 30 and remains almost constant over the entire 0 to 2π phase range. Representative μBLS intensity maps measured for the in-phase and out-of-phase excitation conditions are shown in Fig. 4b. Despite the opposite input phases, nearly identical two-dimensional spin-wave intensity distributions are observed throughout the device. This demonstrates that nonlinear spin-wave propagation redistributes the incoming wave populations such that both phase configurations generate nearly identical effective excitation conditions at the pump region, thereby activating the same nonlinear emission state. Such phase robustness is essential for scalable cascading, as it prevents the accumulation of uncontrolled phase distortions during signal propagation across multiple neuronal stages.

The phase-robust neuronal response originates from nonlinear spin-wave propagation before and within the combining region. Strong nonlinear magnon-magnon interactions redistribute the incoming spin-wave population into nonlinear excitation states that are substantially less sensitive to the externally imposed microwave phases [see Supplementary Materials]. As a consequence, different input-phase configurations evolve toward nearly identical effective excitation conditions at the pump region, allowing the nonlinear activation process to converge into the same regenerated high-emission state. The neuron therefore operates as a nonlinear phase-insensitive element, where a broad range of input phase configurations are transformed into a deterministic and nearly identical output state. Importantly, although nonlinear frequency-shifted spin-wave populations play a central role during propagation and activation, once the neuron switches into the high-emission state, the pump antenna regenerates coherent spin waves at the original microwave excitation frequency. As a result, the frequency-shift information generated during nonlinear propagation is effectively erased from the final output, producing a standardized neuronal output state suitable for cascading.

In addition to phase robustness, the nonlinear magnonic neuron provides intrinsic signal normalization. In our device, spin waves at both the input and output ports are generated through deeply nonlinear excitation. Once the activation threshold is exceeded, the emitted spin-wave intensity is no longer determined by the exact input amplitude or pump power, but instead becomes fixed by the nonlinear excitation condition at the selected operating frequency [34,40]. This self-normalized excitation mechanism produces comparable spin-wave intensities at the input and output ports, thereby enabling direct neuron-to-neuron cascading without additional signal amplification. The time-resolved measurements in Fig. 4c confirm this behavior: the two input signals and the regenerated output signal exhibit comparable intensity levels, while the output displays a slightly delayed rise time. This delay originates from the threshold-triggered activation of the pump antenna by the incoming spin waves rather than from direct microwave excitation.

Based on the nonlinear phase-robust and self-normalized excitation mechanisms discussed above, we experimentally demonstrate deterministic cascading between sequential magnonic neurons. The cascaded structure consists of two interconnected Y-shaped magnonic neurons fabricated from a 47-nm-thick YIG film [37,38] using electron-beam lithography and etching. Five gold stripe antennas were deposited on top of the YIG film: three serving as input antennas and two acting as pump antennas that define the nonlinear activation regions of the two neurons, as illustrated in Fig. 4d, e. The dashed outlines indicate the two neuronal stages, where the output of the first neuron is directly connected to the second neuron through a curved spin-wave waveguide. The microwave powers were adjusted such that the input antennas directly excited propagating spin waves, while the pump antennas operated in the N=2 nonlinear activation regime, where neuronal firing occurs only when spin waves from two inputs arrive simultaneously.

Because of the limited scanning range of the μBLS system, the measurements were performed over 15×7.5 μm$^2$ regions centered on each neuronal stage. Figure 4d shows the cascaded operation when Inputs 1 and 2 of the first neuron are simultaneously activated, while Input 3 of the second neuron is also switched on. The two-dimensional μBLS intensity maps, integrated from 4.6 to 5.5 GHz, reveal a strong spin-wave output from the first neuron corresponding to the activated nonlinear emission state. This regenerated spin-wave signal propagates through the curved waveguide and directly serves as one of the inputs of the second neuron. Together with Input 3, the combined excitation exceeds the activation threshold of the second pump antenna, thereby triggering the second neuron and generating a strong output signal from the second neuronal stage. Figure 4e presents the complementary configuration in which only Input 2 of the first neuron is activated while Input 1 remains switched off. Under this condition, the first neuron remains below its activation threshold and produces only a weak output signal. Consequently, the excitation arriving at the second neuron is insufficient to activate the second nonlinear emission state, and the second neuron also remains in the low-emission state.

These measurements demonstrate deterministic neuron-to-neuron cascading enabled by nonlinear phase-robust signal transfer and intrinsic signal regeneration. Importantly, the regenerated output of the first neuron directly drives the second neuronal stage without external amplification or signal conditioning, establishing a pathway toward scalable integrated magnonic neural circuits.

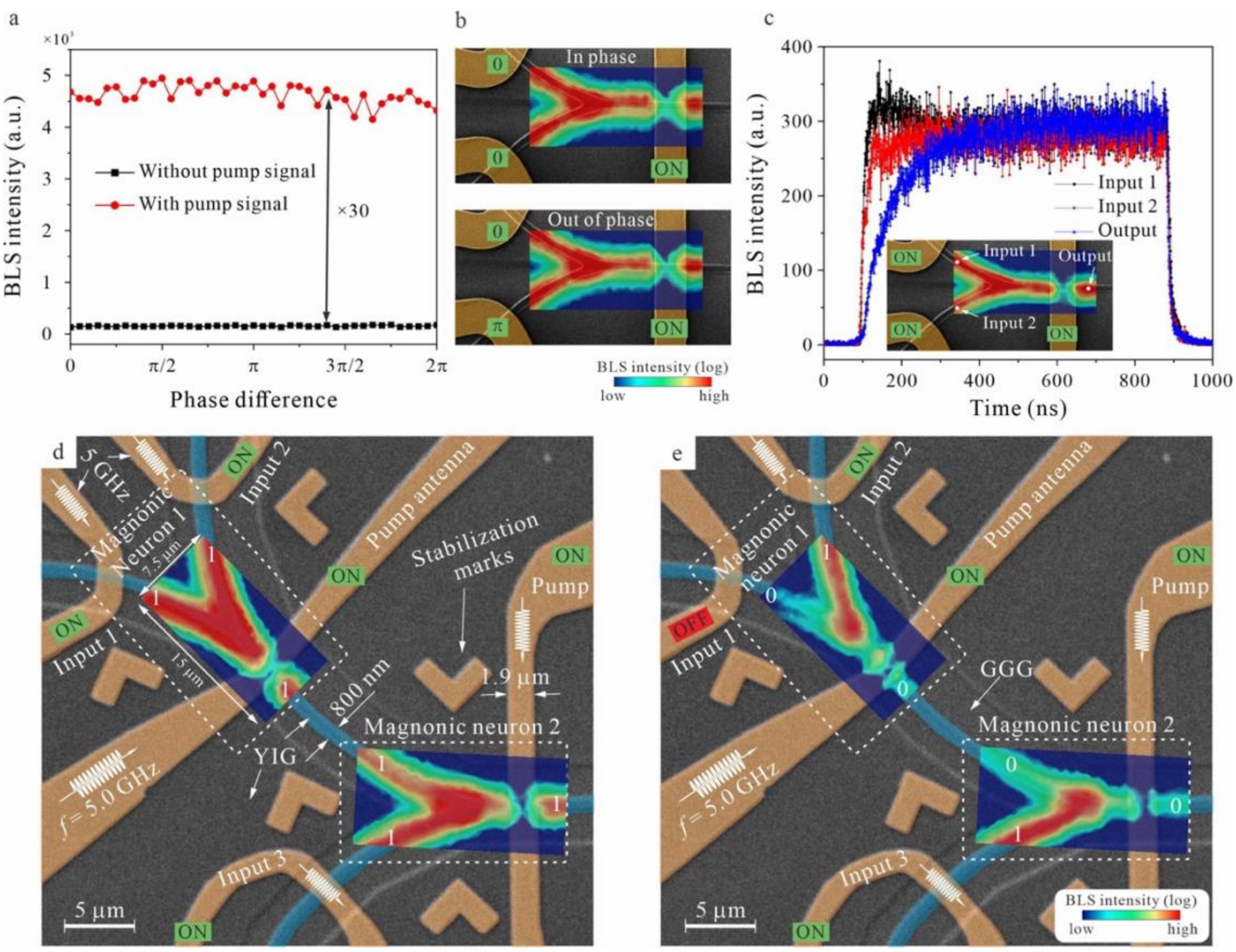

*Figure 4 Phase-robust and cascaded operation of nonlinear magnonic neurons.* ***a***, *Normalized neuronal output intensity as a function of the relative phase difference between the two input spin waves measured with (red) and without (black) pump activation.* ***b***, *Two-dimensional μBLS intensity maps of a simplified Y-shaped magnonic neuron*

*measured for in-phase and out-of-phase excitation conditions. Despite the opposite input phases, nearly identical two-dimensional spin-wave intensity distributions are observed, demonstrating phase-robust nonlinear activation and regeneration of a nearly identical output spin-wave state.* ***c****, Time-resolved measurements of the spin-wave intensities at the two inputs and the neuronal output. The nonlinear excitation mechanism generates a self-normalized output intensity comparable to those of the inputs.* ***d,e****, Deterministic cascading between two interconnected nonlinear magnonic neurons. In* ***d****, Inputs 1 and 2 activate the first neuron, whose output propagates through the curved waveguide and, together with Input 3, activates the second neuron. In* ***e****, only Input 2 is activated, leaving both neurons below threshold. The μBLS intensity maps are integrated from 4.6 to 5.5 GHz. The pump antennas operate in the N=2 activation regime.*

Additional fan-out experiments demonstrating one-to-many distribution of regenerated spin-wave signals are presented in the Supplementary Materials, further confirming the compatibility of the proposed neuron architecture with more complex network topologies.

Building upon the deterministic cascading demonstrated above, we next realize physical pattern recognition in an integrated magnonic neural network composed of seven interconnected threshold neurons. Figure 5a presents the operating principle of a reprogrammable binary pattern-classification network demonstrated for the binary letter patterns “H”, “U”, “S” and “T”. Each letter is encoded using a 3×5 binary pixel matrix corresponding to 15 spatial input channels. For each target letter, characteristic active pixels within the 3×5 binary matrix are selected as the spin-wave input features of the neural network. Pixels corresponding to black regions of the target pattern are assigned to the “on” state, while the remaining pixels are assigned to the “off” state. Simultaneously, the activation thresholds of different neuronal stages are independently programmed through adjustment of the pump powers [details see Supplementary Materials].

As one representative example, recognition of the letter “H” is implemented by selecting a specific subset of active pixels from the 3×5 binary input matrix and mapping them onto the physical input channels of the integrated magnonic neural network. The activation thresholds of the seven neurons are independently programmed through the pump powers, as indicated in Fig. 5a. Under this configuration, only input patterns matching the selected features can successfully propagate through all neuronal stages and generate a high-output state. Because the network operation is governed by programmable threshold activation rather than fixed interference geometries, the same physical architecture can be dynamically reconfigured to classify different binary patterns.

Figure 5b presents an optical image of the fabricated device. The central region contains the integrated magnonic neural network, while a total of 17 microwave antennas, including both input and pump antennas, are connected through wire bonding for electrical excitation and control. Figure 5c shows a SEM image of the central network region. The yellow structures correspond to the input antennas, the red structures to the pump antennas, and the blue regions to the YIG waveguides. The integrated neural circuit consists of two three-input nonlinear neurons and five two-input nonlinear neurons arranged in a four-layer hierarchical architecture, enabling multi-stage nonlinear processing of spin-wave signals.

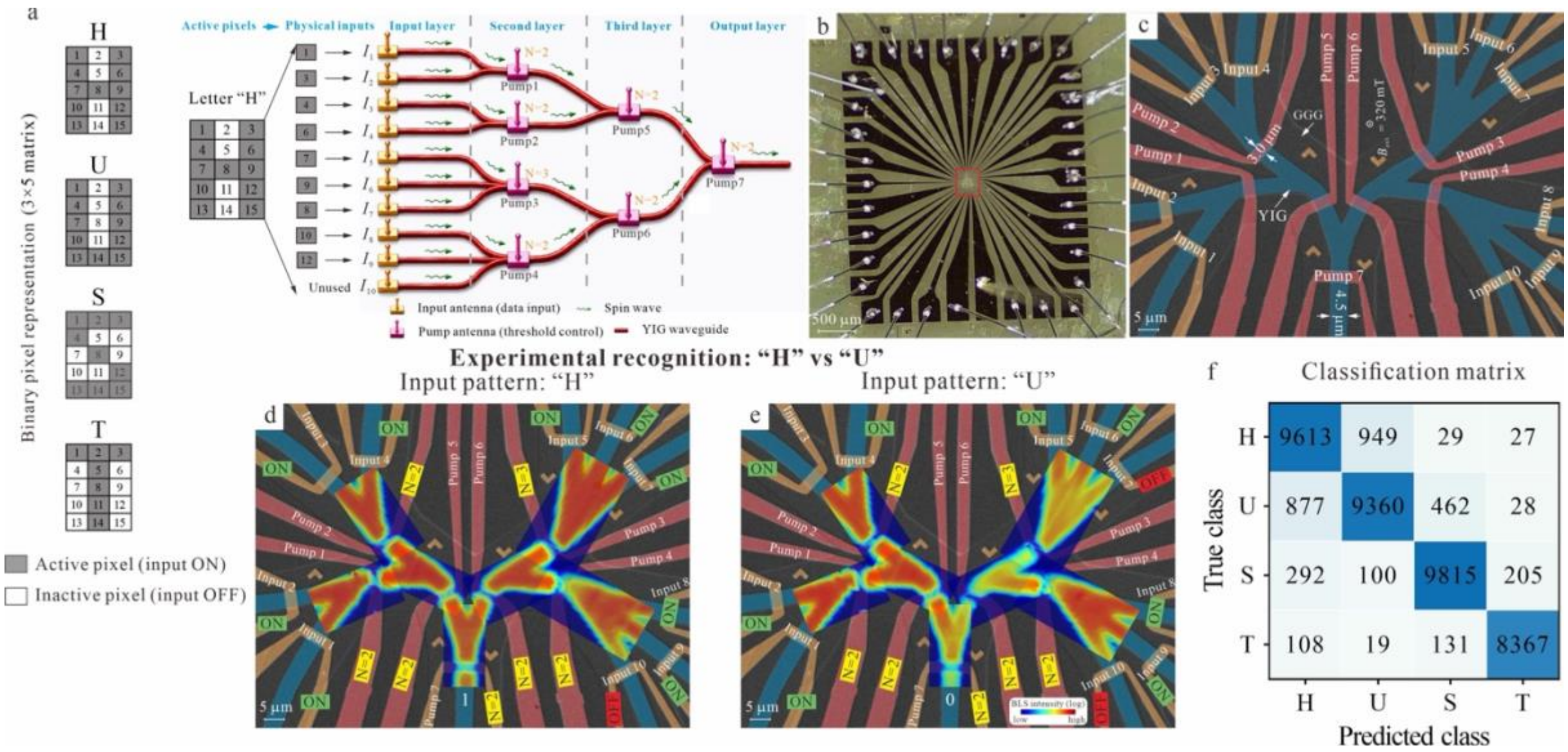

*__Figure 5 Physical pattern recognition using an integrated magnonic neural network. a,__ Schematic illustration of the integrated magnonic neural network designed for classification of the binary letter patterns "H", "U", "S" and "T". Each letter is encoded using a 3×5 binary pixel matrix, whose active pixels are mapped onto the physical spin-wave input channels of the neural network. For each target pattern, specific feature inputs and pump-threshold configurations are selected. As one example, recognition of the letter "H" is implemented using the feature pixel 1, 3, 4, 6, 7, 8, 9, 10 and 12, while the activation thresholds of the seven neurons are independently programmed as indicated in the network schematic. __b,__ Optical image of the fabricated magnonic neural chip. The central region (red rectangular region) contains the integrated magnonic neural network, while a total of 17 microwave antennas, including input and pump antennas, are connected through wire bonding for electrical excitation and control. __c,__ SEM image of the central network region. Yellow regions correspond to the input antennas, the red structures to the pump antennas, and blue regions represent the YIG waveguides. The integrated neural circuit consists of two three-input neurons and five two-input neurons arranged in a four-layer architecture, enabling multi-stage nonlinear processing and cascading of spin-wave signals. __d,e,__ Experimental μBLS intensity maps measured using the input and pump-threshold configuration optimized for recognition of the letter "H". In __d__, the input pattern corresponding to "H" successfully activates the cascaded neuronal network, producing a strong output signal. In __e__, the input pattern corresponding to "U" fails to activate the top right corner neuron because pixel 8 (input 7) is inactive, interrupting the cascading process and suppressing the final output. Due to the limited scanning area of the μBLS system, the intensity maps were assembled from multiple spatial scans, resulting in slight overlaps between adjacent regions. __f,__ 4×4 classification matrix obtained using different input and pump-threshold configurations for classification of the binary patterns "H", "U", "S" and "T". The values represent the integrated spin-wave intensities measured at the output waveguide by scanning the μBLS laser spot across the waveguide width. Strong responses are observed along the diagonal elements, demonstrating successful physical pattern recognition in the integrated magnonic neural network.*

Figures 5d and 5e present the experimental pattern-recognition results measured by μBLS. Figure 5d shows the response of the integrated network using the input and pump-threshold configuration programmed for classification of the letter "H". When the binary pattern corresponding to "H" is applied, the selected spin-wave inputs collectively satisfy the activation conditions of all neuronal stages, resulting in strong spin-wave emission at the output waveguide corresponding to the neuronal high-output state ("1"). In contrast, the same network configuration fails to recognize the letter "U", as shown in Fig. 5e. The critical difference between the two input patterns is that pixel 8 (Input 7) is inactive in the "U" pattern. As a result, the top right corner neuron remains below its activation threshold, interrupting the cascaded signal propagation and suppressing the final output. The network therefore remains in the low-emission state ("0"). Corresponding micromagnetic simulations reproduce the same network operation and classification behavior, as presented in the Supplementary Materials.

The overall classification performance is summarized in the 4×4 classification matrix shown in Fig. 5f. For each measurement, the input and pump-threshold configuration optimized for one target letter was used to classify all four binary patterns "H", "U", "S" and "T". The values in the matrix represent the integrated spin-wave intensity measured at the output waveguide. Experimentally, the spin-wave intensity was obtained by scanning the μBLS laser spot across the width of the output waveguide and integrating the detected magnon intensity. The diagonal elements of the matrix are significantly larger than the off-diagonal elements, demonstrating successful physical classification of the four binary letter patterns. Although the patterns "H" and "U" differ by only a single input feature, the corresponding output intensities still differ by more than one order of magnitude, highlighting the strong nonlinear selectivity of the integrated magnonic neural network.

**Discussion**

Our work establishes a scalable physical platform for integrated wave-based neural computing using nonlinear magnons. By combining programmable weighting, tunable threshold activation, intrinsic signal regeneration and deterministic cascading within a single nanoscale architecture, the demonstrated magnonic neurons overcome several longstanding obstacles that have limited the scalability of wave-based information processing systems. In conventional wave-computing architectures, computation is typically performed through interference of propagating waves, making the output highly sensitive to accumulated phase errors, propagation losses and fabrication-induced variations. As a result, reliable multi-stage integration generally requires external amplification, calibration or active error compensation. Here, nonlinear spin-wave dynamics transform diverse input states into robust regenerated output states, enabling deterministic neuron-to-neuron communication without external signal restoration, synchronization or signal conditioning.

More broadly, the demonstrated neuron architecture introduces a different paradigm for wave-based information processing. Rather than encoding computation solely in interference patterns, the present approach exploits nonlinear collective dynamics to perform thresholding, normalization and signal regeneration directly within the physical medium. The resulting neurons operate as local decision-making units whose activation depends on the collective state of incoming waves rather than on precise phase relationships. This mechanism provides a route toward scalable wave-native neural hardware, where complex computational functionality emerges from interactions among interconnected nonlinear wave neurons.

The significance of this approach extends beyond magnonics. Threshold activation, signal regeneration and cascading are fundamental requirements for large-scale neural computation irrespective of the underlying physical platform. The present results demonstrate that these functionalities can emerge directly from nonlinear wave dynamics without requiring auxiliary electronic circuitry. In contrast to photonic systems, where nonlinear activation often relies on additional materials, resonant structures or electronic feedback, magnons naturally combine nanoscale wavelengths with strong intrinsic nonlinear interactions, enabling neuronal functionalities to be realized within a compact wave medium. The demonstrated architecture therefore establishes nonlinear collective excitations as functional neural building blocks and suggests a general pathway toward scalable physical neural networks based on interacting waves.

Although the present demonstration contains seven interconnected neurons, the current network size is primarily limited by planar microwave routing and electrical interconnection density rather than by the neuron architecture itself. Because threshold activation, signal regeneration and phase stabilization are performed locally within each neuron, increasing network complexity does not inherently require global phase control or external restoration stages. Additional fan-out experiments presented in the Supplementary Materials further demonstrate one-to-many distribution of regenerated neuronal signals, supporting the compatibility of the proposed architecture with larger-scale neural-network topologies. Together, these results demonstrate that threshold activation, signal regeneration, fan-out and deterministic cascading can emerge directly from nonlinear collective wave dynamics, providing a foundation for scalable wave-native neural hardware beyond conventional charge-based electronics.

**Method**

**YIG Nanoscale structures fabrication**

The 47-nm- and 103-nm-thick YIG films were grown on 500-μm-thick (111)-oriented gadolinium gallium garnet (GGG) substrates by liquid-phase epitaxy (LPE) [37,38]. The material parameters of the unpatterned films were characterized using stripline vector network analyser ferromagnetic resonance and BLS spectroscopy, yielding a saturation magnetization of $M_s = (140.7 \pm 2.8)$ kA/m, a Gilbert damping parameter of $\alpha = (2 \pm 0.08) \times 10^{-4}$, an inhomogeneous linewidth broadening of $\mu_0 \Delta H_0 = (0.18 \pm 0.01)$ mT, and an exchange constant of $A_{\mathrm{ex}} = (4.22 \pm 0.21)$ pJ/m. These values are representative of high-quality thin-film YIG [37,38,48].

The individual Y-shaped magnonic neuron (Fig. 4b) were fabricated from a 103-nm-thick YIG film using a Cr/Ti hard mask and ion-beam milling, following the procedure described in Ref. [48]. The three-input as shown in Fig. 1c and Fig. 2, the cascaded Y-shaped neurons as shown in Fig. 4d and integrated neuron circuit as shown in Fig. 5c were fabricated from a 47-nm-thick YIG film, where a positive electron-beam resist (AR-P 6200.09) was used directly as the etching mask during ion-beam milling. Because a positive-tone resist process was employed, only the device perimeter needed to be exposed, significantly reducing electron-beam lithography time. As a result, during the subsequent ion-beam etching, the YIG was completely removed only in the vicinity of the patterned devices, while the surrounding regions remained covered by the unetched YIG film. The stripe antennas with widths of 2-3 μm, were patterned using standard electron-beam lithography. It consists of ~ 100 nm thick gold and 10 nm thick titanium layer (for adhesion).

**BLS measurements**

A single-frequency laser with a wavelength of 457 nm was focused onto the sample using a 100× microscope objective with a numerical aperture of NA = 0.8. The laser power at the sample was approximately 3 mW. A uniform out-of-plane magnetic field of 320 mT was applied using a 70-mm-diameter NdFeB permanent magnet. Microwave signals with various powers and frequencies were supplied to the antennas to directly excite or amplify spin waves.

For two-dimensional BLS mapping, the sample was scanned relative to the fixed laser spot in steps of a few hundred nanometres using a piezoelectric positioning stage. Stabilization marks fabricated on the sample surface ensured precise realignment of the laser spot, allowing consistent positioning throughout the long measurement cycles.

**Micromagnetic simulations**

The micromagnetic simulations were performed using the GPU-accelerated package MuMax3, incorporating both exchange and dipolar interactions to compute the full space- and time-dependent magnetization dynamics of the investigated structures [49]. Material parameters corresponding to nanometre-thick YIG films were used [37,38,48]: a saturation magnetization of $M_s = 1.407 \times 10^5$ A/m and an exchange constant of $A =$ 4.2 pJ/m. The Gilbert damping parameter was set to $\alpha = 7 \times 10^{-4}$ to effectively account for the inhomogeneous linewidth, which cannot be included directly in MuMax3. To suppress spin-wave reflections at the device boundaries, the damping was exponentially increased to 0.5 near the ends of the structure.

The simulation mesh size was defined as $20 \times 20 \times 103\ \text{nm}^3$ (with a single cell along the thickness) for the YIG waveguide. An external magnetic field of $B_{\text{ext}} = 320$ mT was applied along the out-of-plane direction, providing full saturation of the film [49].

**Data availability**

The data that support the plots presented in this paper are available from the corresponding authors upon reasonable request.

**Code availability**

The code used to analyse the data and the related simulation files are available from the corresponding author upon reasonable request.

**Acknowledgements**

This work was supported from the National Key Research and Development Program of China (Grant No. 2023YFA1406600) and the National Natural Science Foundation of China (Grant No. 12574118). R. V. acknowledge support by the NAS of Ukraine, project #0124U000270. P. P. acknowledges support by the Deutsche Forschungsgemeinschaft (DFG, German Research Foundation) -TRR 173–268565370 ("Spin + X", Project B01) and by the European Research Council within the Starting Grant No. 101042439 "CoSpiN". A. V. C. acknowledges the financial support of the Austrian Science Fund (FWF) by means of grant MagNeuro no. 10.55776/PIN1434524. C. D. acknowledges the financial support by the Deutsche Forschungsgemeinschaft (DFG, German Research Foundation) - 271741898.

**Author Contributions**

M. G. performed BLS measurements with help from X. G., Z. Z.. X. J. and K. D. fabricate the nanoscale sample with help from C. G., Y. R., K. C. and J. L.. X. J. and M. G. performed micromagnetic simulations. R. V. provided theoretical support and analysis. C. D. grew the YIG films. A. C., P. P., and Q. W. led this project. Q. W. conceived the idea and wrote the manuscript with the help of all the coauthors. All authors contributed to the scientific discussion and commented on the manuscript.

**Competing Interests**

The authors declare no competing interests.

**Correspondence** and requests for materials should be addressed to Q. W.